\documentstyle[epsf]{ioplppt}

\begin{document} 
\JPA

\title{Operator Spectrum and Exact Exponents of the Fully Packed Loop
Model}

\author{Jan\'e Kondev}
\address{Physics Department, Brown University, Providence, Rhode Island 
02912-1843}
\author{Jan de Gier and Bernard Nienhuis}
\address{Instituut voor Theoretische Fysica, Universiteit van Amsterdam,
Valckenierstraat~65, 1018 XE Amsterdam, The Netherlands}

\begin{abstract}
We develop a  Coulomb gas description of the critical fluctuations  in 
the fully packed loop model on the honeycomb lattice. 
We identify  the complete operator spectrum of this model in terms of 
electric and magnetic {\em vector}-charges,  and we calculate the  
scaling dimensions of these operators exactly. We also study the  
geometrical properties
of loops in this model, and we derive exact results for 
the fractal dimension and the loop size distribution
function. A review of the many different representations of this model that 
have recently appeared in the literature, is given. 
\end{abstract}

%\maketitle

%%%%%%%%%%%%%%%%%%%%%%%   MACROS    %%%%%%%%%%%%%%%%%%%%%%%%%%%%%%%%%%%%%%%%

\newcommand{\be}[1]{\begin{equation}\label{#1}}
\newcommand{\ee}{\end{equation}}
\newcommand{\bea}[1]{\begin{eqnarray}\label{#1}}
\newcommand{\eea}{\end{eqnarray}}
\newcommand{\eqref}[1]{equation~\eref{#1}}
\newcommand{\Rs}{\sf I\hskip-2pt R}              
\newcommand{\Zs}{\mbox{\sf Z\hskip-5pt Z}}         
\newcommand{\Cs}{\rm C\!\!\!I\:}

%%%%%%%%%%%%%%%%%%%%%%%%%%%%%%%%%%%%%%%%%%%%%%%%%%%%%%%%%%%%%%%%%%%%%%%%%%%%

\section{Introduction}

Loop models, loosely speaking, are statistical models which have as basic
building blocks loops that run along the bonds of a two-dimensional lattice. 
In order to define a loop  model one assigns Boltzmann weights to the  
different loop configurations. This is usually implemented
by assigning weights to the 
different vertex configurations allowed by the loop model, and a weight 
to the loop as a whole; this loop-weight is usually referred to as 
the fugacity.

Loop models have  attracted attention 
recently as representations of certain exactly-solvable vertex models
which can be used to construct restricted solid-on-solid models\cite{Xpasq1}, 
some of which admit an off-critical extension \cite{Xwarnaar}.
They are also particularly simple examples of models that allow
for Monte-Carlo simulations with non-local loop updates,
which have been recently studied 
as algorithms that reduce critical slowing down \cite{Xevertz}. 
In a completely different setting, 
loop models appear as space-time diagrams in the path integral representation 
of one-dimensional quantum spin chains, and many quantities defined in terms
of  the 
spins can be reexpressed in the language of loops \cite{Xbruno}. 
For instance, it
can be shown that the spin-spin correlation function in the 
antiferromagnetic Heisenberg spin 
chain, is 
given by the loop correlation 
function in the appropriate loop model \cite{Xbruno}. The loop correlation 
function  measures the probability that two points on the lattice 
belong to the same loop. 

In two-dimensional classical 
spin models loops are typically encountered as domain
boundaries, e.g., Bloch walls in the Ising model, or as graphical
representations of high-temperature expansions. 
Recently, Cardy \cite{Xcardy} has
calculated different geometrical properties of cluster boundaries in the 
$O(n)$ model on the honeycomb lattice, using the loop representation 
of this model.

In this paper we study the fully packed loop (FPL) model on the honeycomb
lattice. The FPL model was introduced by Reshetikhin \cite{Xresh}, 
and independently by Bl\"ote 
and Nienhuis \cite{Xblote}, as the zero-temperature 
limit of the $O(n)$ model. 
Using numerical transfer-matrix methods, Bl\"ote and Nienhuis were able to 
to show that this loop model defines a universality class distinct from 
the previously studied low temperature phase of the $O(n)$. Exact values of
the critical  exponents and the conformal charge of the FPL model  were 
subsequently determined by Batchelor {\em et al.} \cite{Xbatch}, who found a 
Bethe ansatz solution of  the model.        

In the FPL model
nonintersecting loops are placed along the bonds of a honeycomb lattice
so that {\em each} vertex of the lattice is covered by a loop.
The partition function is given by
\begin{equation}
\label{part}
Z_{FPL} = \sum_{\cal G} \: n^{N({\cal G})}
\end{equation}
where $N$ is the number of loops in the fully packed configuration 
${\cal G}$, and $n$ is the loop fugacity.
 
This model undergoes a phase transition as a function of the loop fugacity
$n$. For values of $n$ approaching zero, configurations with a small number
of big loops are favored; in the limit $n\rightarrow 0$ a single loop covers 
the whole lattice.\footnote{This limit was studied by Batchelor {\em et al.} 
\cite{Xbatch} who obtained exact results for Hamiltonian walks on the 
honeycomb lattice.}
Loops of all sizes will be present on the lattice as we increase $n$. 
This is equivalent to having a diverging correlation length
in the system \cite{Xkast}, and the model
is critical with power law correlations. At $n=2$ the FPL model undergoes
a Kosterlitz-Thouless type of transition \cite{Xresh,XKH} into a long-range
ordered state, in which there exists a largest loop on the 
lattice which is roughly the size of the correlation length.
In the $n\rightarrow \infty$ limit the fully packed loop model is perfectly 
ordered, with all the loops having the minimal length of six, and occupying 
one of the three sublattices of hexagonal plaquettes. 

Here we focus our attention on 
the FPL model along the critical line 
($0 \leq n \leq 2$), which
was also the focus of the above mentioned numerical transfer-matrix 
study, and of the Bethe ansatz solution. 
Using a nested Bethe ansatz Batchelor {\em et al.} \cite{Xbatch} calculated
the scaling  dimensions  of the 
``watermelon" operators along the critical line. 
The watermelon correlation function 
is defined as the  probability that $m$ loop segments   
meet in the neighborhood of two points separated by $\vec{ r}$ \cite{Xdupl}. 
Here we rederive the same results from a 
Coulomb gas approach in which the loop model is mapped to an interface 
model. In the interface representation  ``watermelon''
scaling dimensions become associated with vortices whose 
topological charges are vectors in the triangular lattice. 
Furthermore, this approach
allows us to identify the {\em complete}  operator spectrum of
the FPL model and make contact with known results from conformal field 
theory. In particular, we calculate exactly 
the temperature dimension  found numerically
by Bl\"ote and Nienhuis \cite{Xblote}, that  does not  appear in the 
Bethe Ansatz solution of Batchelor {\em et al.} \cite{Xbatch}. We also 
identify defect configurations in the loop model that generalize the 
``watermelon'' configurations, and we calculate the critical exponents 
associated with them.

The main shortcoming of 
the Coulomb gas approach, in general, 
is that it usually relies on some exact information about 
the model which can then be used to calculate the  
value of the renormalized coupling \cite{Xnienrev}. Once the coupling 
is known all the exponents can be calculated exactly.
We will show
that in the FPL model the coupling can be determined exactly by 
identifying the marginal operators in this model. 
The existence of these operators is required by certain consistency 
conditions placed 
on the conformal field theory which describes the scaling limit of a lattice
model; this was discussed at length by Dotsenko and Fateev \cite{Xdots}.

Many of our results for the critical exponents of the FPL model 
have been found previously from a Bethe Ansatz solution \cite{Xbatch}. 
Our main motivation for pursuing the Coulomb gas approach 
is its relative simplicity, and 
the geometrical interpretation of the operator spectrum of the FPL model,
which it offers. Furthermore, this approach  allows us to identify 
the conformal field theory that describes the scaling limit of the FPL 
model, which can then be used to study the critical 
properties of this model in detail, 
using the many conformal techniques at our disposal. 

This paper is organized as follows:
in \sref{reps} we have collected the 
different known representations of the FPL model, and mappings from 
one to the other are made explicit.
In \sref{effth} 
we introduce an effective field theory of the FPL model, and
\sref{theCG} is devoted to the construction of
the associated Coulomb gas and the 
calculation of the scaling dimensions of different operators. 
These results are used in \sref{geom} to calculate the geometrical 
exponents for loops in the FPL model.

\section{Representations of the FPL model}
\label{reps}

The FPL model has many different representations, some of which 
have been independently studied.
Here we review the mappings between the different representations.

The FPL model for $0\leq n \leq 2$ is equivalent to the three-colouring
model on the 
honeycomb lattice introduced by Baxter \cite{Xbaxter}. The three-colouring
model is defined by colouring the bonds of the honeycomb lattice with 
three different colours, say $A$, $B$, and $C$, in such a way that no two 
bonds of equal colour meet at a vertex.\footnote{These type of graph
colourings are known in the mathematics literature as {\em edge colourings}.} 
 For $n=2$ each colouring is
given equal statistical weight. If we choose any two colours, say $B$ and $C$,
then the bonds coloured with these two colours form a fully packed loop 
configuration on the honeycomb lattice.  Each loop can be coloured with 
alternating
colours $B$ and $C$ in two ways ($B$-$C$-$B\ldots$ or $C$-$B$-$C\ldots$), 
and is therefore assigned a fugacity $n=2$. The FPL model 
away from the $n=2$ point can also be mapped to a colouring model, but now
the weights of the different colourings will have to be modified; 
see \sref{n<2}. 

If we consider the three colours as Potts spins placed at the centers of the 
bonds of the honeycomb lattice, then the three-colouring model 
describes the ground state of the 
three-state antiferromagnetic 
Potts model on the Kagom\'e lattice studied by Huse and Rutenberg
\cite{Xhuse}. The Potts model is defined by the Hamiltonian (energy functional)
\begin{equation}
\label{kagome}
    H = |J| \sum_{<ij>} \delta_{\sigma_i,\sigma_j} \; ,
\end{equation}
where the sum goes over nearest-neighbors, and the spins $\sigma_i$ live
on the vertices of the Kagom\'e lattice. At zero temperature the only 
allowed states are ones where on every triangular plaquette all three 
spins are present. This ground state manifold is critical in the sense that 
correlation functions of the spins decay with distance as 
power laws \cite{Xhuse}. 
 
\begin{figure}[tb]
\epsfxsize=7cm \epsfbox{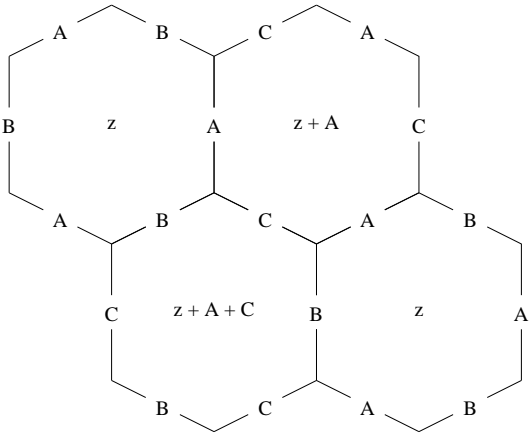}
\caption
{
\label{3-ideal-state}
One of six symmetry related ideal states
of the three-colouring model. In an ideal state all the plaquettes
are coloured with two colours only. The microscopic heights $\vec{z}$ 
are defined at the centers of the plaquettes, and the change in $\vec{z}$, when
going from one plaquette to the neighboring one,  is
determined by the colour of the crossed bond. The ideal state is 
macroscopically
flat, in the sense that the variance of the microscopic height is minimal.
}
\end{figure}

The three colouring model can be mapped to a solid-on-solid model 
that describes a two dimensional interface 
in {\em four} spatial dimensions \cite{Xhuse,XKH}. 
This is accomplished by placing a 
two component  microscopic height  
$\vec{z}$ at  the 
center of each plaquette of the honeycomb lattice; see \fref{3-ideal-state}. 
The change in $\vec{z}$ when going from a plaquette to
the neighboring one is given by $\vec{A}$, $\vec{B}$, 
or $\vec{C}$, depending on
the colour of the bond that is crossed; 
the vectors $\vec{A}$, $\vec{B}$, and  $\vec{C}$ point to the vertices
of an equilateral triangle,
\begin{equation}
\label{vecs}
\vec{A} = (\frac{1}{\sqrt{3}}, 0) \; , \; \; 
\vec{B} = (-\frac{1}{2 \sqrt{3}}, \frac{1}{2}) \; , \; \;
\vec{C} = (-\frac{1}{2 \sqrt{3}}, -\frac{1}{2}) \; ,
\end{equation}
where we have chosen the normalization for later convenience.
We also  adopt the convention that the microscopic height {\em increases}
by $\vec{A}$, $\vec{B}$, or  $\vec{C}$ when going clockwise around an up 
pointing triangle of the dual lattice (i.e., around a vertex of the honeycomb 
lattice in the shape of the letter Y); see \fref{3-ideal-state}.
Up to an arbitrary choice of a single height, say  at  the origin, 
the mapping of the colouring to the heights is one-to-one.
Each allowed height configuration is given equal statistical weight.

%---------------------------------------------------------------------------

%---------------------------------------------------------------------------

It has been recently shown by Di Francesco and Guitter \cite{Xdifrancesco}
that the three colouring model can be mapped to
a folding model of the triangular lattice, introduced by Kantor and
Jari\'c \cite{Xkantor}. The allowed configurations in the
folding model are given by all the possible {\em complete} foldings of the
triangular lattice.\footnote{This model is a special case of a more 
general folding model due to Shender {\em et al.} 
\cite{Xshender}, which is equivalent to the 
ground states
of the antiferromagnetic {\em Heisenberg model} on the Kagom\'e lattice.} 
A folding configuration can be 
specified by giving the direction ($+$ or $-$) of the normal to 
each elementary triangle 
in the folded state. Now if we place the three colours, $A$, $B$, and $C$,
 on the bonds of every
elementary triangle  in a clockwise ($+$) or anticlockwise ($-$) 
fashion we obtain a three colouring configuration
of the dual honeycomb lattice. This is a six-to-one mapping, since for a
given folded configuration one is free to choose one of six 
colour configurations
around a single triangle, which then fixes all the rest.

\section{Effective-field theory}
\label{effth}

In this section we propose an effective field theory for 
the long wavelength fluctuations of the interface model, which is one
of the representations of the FPL model discussed in the previous 
section. Here we focus on the $n=2$ case which is equivalent to the 
three-colouring model with equal statistical weight for all the colourings, 
and extend to $n<2$ in the following section. 

We motivate the long-wavelength theory of the interface model 
by a coarse-graining procedure of the microscopic heights $\vec{z}$, which
is implemented as follows:\footnote{Details of the height construction 
for the $n=2$ FPL model, as well as for other critical ground states,
can be found in \cite{XKHNPB}.}
First, we define the {\em ideal states} which we use to coarse-grain the
three-colouring model. Ideal states
are edge colouring states in which every elementary plaquette of the honeycomb
lattice is coloured with two colours only, \fref{3-ideal-state}. 
These states are flat, in the sense that they have the smallest variance
of the microscopic height, and we argue that the
free energy of the colouring model
\footnote{The free energy of the 
three-colouring model is purely entropic in origin, in the sense that 
the partition function is simply equal to the number of different edge 
colourings.}
is dominated by fluctuations around the ideal states.
Namely, the smallest change on the lattice, that is allowed by 
the constraints of the three-colouring model, is an exchange of colours 
along a {\em loop} of alternating colour (eg. $C$-$B$-$C\ldots$ 
to $B$-$C$-$B\ldots$). The 
ideal states {\em maximize} the number of loops that allow for these loop 
exchanges,
and it is this property that selects them out. This entropic selection effect
is close in spirit to the ``order by disorder" effect introduced by
Villain \cite{Xvill}. In the FPL model ideal states are the ones 
selected in the $n\rightarrow \infty$ limit, while in the folding model these
are states where the  
triangular lattice has been folded to a single triangle.  

Second, we divide the honeycomb lattice into domains so that each domain
represents a different ideal (flat) state.
To each domain 
we assign a {\em coarse grained height} $\vec{h}$, which is equal to
the microscopic height averaged over the domain; 
$ \vec{h} = \langle \: \vec{z} \: \rangle$.

\begin{figure}[tb]
\epsfxsize=7cm \epsfbox{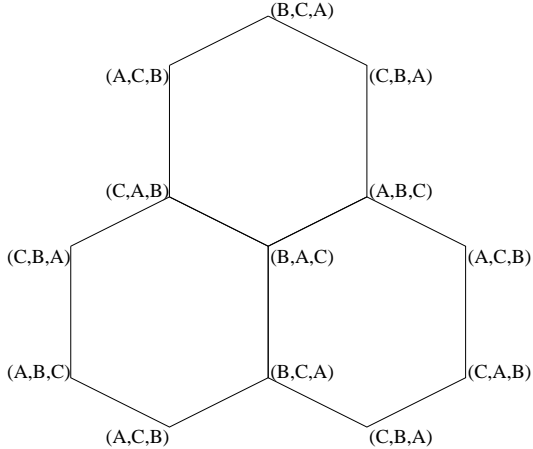}
\caption[The ideal state graph of the three-colouring model.]
{
\label{3-ideal}
The ideal state graph of the three colouring model is a honeycomb lattice in
height space:
each vertex is associated with a particular ideal state, and the six
different ideal states form a hexagonal plaquette. The 
ideal states are labeled by the colour configuration  
$(\sigma_1,\sigma_2,\sigma_3)$ 
of the bonds around a common vertex of the coloured  lattice. 
The vertices in the ideal state graph that 
correspond to the {\em same} ideal state (say $(C,B,A)$) form a triangular 
lattice which is the {\em repeat} lattice of the three-colouring model.    
}
\end{figure}

The coarse grained heights associated with the six different ideal states 
(one for every permutation of the three colours) form a  
honeycomb lattice which we call the {\em ideal state graph} ${\cal I}$; 
see \fref{3-ideal}.
The side of the elementary hexagon of ${\cal I}$ 
is $1/3$, in the units chosen for the
vectors representing the colours, \eqref{vecs}. Nodes of ${\cal I}$ that 
correspond to the {\em same} ideal state form a triangular lattice 
with an elementary triangle of side $1$. This lattice we call the 
{\em repeat} lattice ${\cal R}$; in the following section we will show
that  vectors $\vec{b}\in {\cal R}$ are the 
magnetic vector-charges in the Coulomb gas associated with the FPL model.

%---------------------------------------------------------------------------

%---------------------------------------------------------------------------

Finally, we consider the continuum limit of the interface model in which the 
discrete heights, defined over different ideal state domains, are replaced 
with a continuously varying height field 
$\vec{h}(\vec{r}) \equiv (h_1(\vec{r}), h_2(\vec{r}))$. The dimensionless 
free energy $f$ of the interface, which is entropic in origin, is assumed
to be of the form
\begin{equation}
\label{free1}
f = \int \d^{2}\vec{r} \: \left[\pi g (|\nabla h_1|^2 + |\nabla h_2|^2) 
                         + V(\vec{h}) \right] \; , 
\end{equation}
where $2\pi g$ is usually referred to as the {\em stiffness}. 
$V(h)$ is a periodic potential with the periodicity given by the 
ideal state graph, i.e.,
\begin{equation}
\label{periodicity}
     V({\bf h} + {\cal I}) = V({\bf h}) \; .
\end{equation}

The free energy $f$  defines an  effective field theory of
the three-colouring model; the assumption being made is that it
correctly describes the long-wavelength fluctuations of the
microscopic height $\vec{z}$.
The periodic potential $V(\vec{h})$, which is usually referred to as the
{\em locking potential}, favors the heights to take
their values on ${\cal I}$, while the first term is the contribution
to the free energy due to  fluctuations
around the flat ideal states.
Therefore, the assumption that the
effective field theory of the three-colouring model is given by
\eqref{free1} is directly
related  to the intuitive idea put forward earlier,
that the free energy of the
three-colouring model is dominated by fluctuations 
around the ideal states.

The locking potential
is periodic with the periodicity of ${\cal I}$. Thus, the
three-colouring  model, in its interface representation,
undergoes a roughening transition for some value of the coupling
$g = g_r$~\cite{Xforgacs}. If the coupling $g$ satisfies $g < g_r$, then
the locking potential  in \eqref{free1} becomes irrelevant, in the
renormalization group sense, and the three-colouring model is described by
a Gaussian model with a free energy
\begin{equation}
\label{freegauss}
 f = \pi g  \int \d^2 \vec{r} \: (|\nabla h_1|^2 + |\nabla h_2|^2) \; .
\end{equation}
In the case that the
locking potential is relevant ($g>g_r$),
the three-colouring model will lock into long range
order in one of the ideal states. This would imply a finite correlation 
length in the FPL model, roughly the size of the largest loop in the system. 
We will see in the following section that for the
the three-colouring model  which is equivalent to the $n=2$ FPL model,
$g$ is {\em equal} to  $g_r$, so the interface is {\em at} the
roughening transition (i.e., the locking potential is {\em marginal}). 
For values of the fugacity $n<2$ it will be shown that
$g<g_r$, but due to the presence of the background charge the the locking
potential remains marginal.\footnote{Marginal roughness seems to be 
a general property of critical loop models in two dimensions \cite{XJKunp}.}

\section{Coulomb Gas}
\label{theCG}

Here we develop the Coulomb gas description of the FPL model based on 
its interface representation. We determine the spectrum of  
possible electric and magnetic charges and we calculate the 
scaling dimensions of operators associated with them.
These are compared to recent 
numerical results \cite{Xblote},
and to results from a Bethe Ansatz solution of the FPL model \cite{Xbatch}.  
We first consider the $n=2$ FPL model which is described by a simple
Gaussian field theory, \eqref{freegauss}. 
The $n<2$ case is treated by perturbing the 
$n=2$ theory with an integrably marginal operator \cite{Xchoud}, 
and introducing a background charge in the Coulomb gas \cite{Xdots}.

\subsection{FPL  model at $n=2$}

In constructing the effective field theory of the three-colouring model,
which maps to the $n=2$ FPL model,  
we found that the height field 
$\vec{h}$ is compactified on the triangular lattice $\cal{R}$.
Therefore, any local lattice operator $\Phi(\vec{r}\,)$
uniform in the ideal states is periodic in height space, and it can 
be written as a Fourier series
\begin{equation}
\label{fourier}
      \Phi(\vec{r}\,) = \sum_{{\vec{G}} \in \cal{R}^\ast} \Phi_{{\vec{G}}}
     \: \e^{\i 2 \pi {\vec{G}} \cdot  {\vec{h}}({\vec{r}})} \; .
\end{equation}
Here $\cal{R}^\ast$ is the lattice dual to $\cal{R}$, i.e., 
$\vec{G}\cdot\vec{b}\in \Zs$ for any two vector-charges 
$\vec{b}\in\cal{R}$ and  $\vec{G}\in\cal{R}^\ast$. 
The Gaussian field theory, \eqref{freegauss}, describes the vacuum phase of a
two-dimensional Coulomb gas of electric ($\vec{G}$) 
and magnetic ($\vec{b}$) vector-charges \cite{Xnienrev}.

The scaling dimension of $\Phi(\vec{r}\,)$ is equal to the scaling 
dimension of the most relevant {\em vertex} operator
$\exp{(\i 2 \pi {\vec{G}} \cdot  {\vec{h}}({\vec{r}}))}$
appearing in its Fourier expansion. The scaling dimension $x(\vec{G})$, 
of a  vertex operator, can be easily determined from its two-point correlation
function, 
\be{vertexcor}
    \langle \e^{\i 2 \pi {\vec{G}} \cdot {\vec{h}}({\vec{r}})} 
    \e^{-\i 2 \pi \vec{G} \cdot \vec{h}(0)} \rangle \sim r^{-2 x(\vec{G})} \; .
\ee
Namely, using a general property of a Gaussian distributed random field 
\be{vertexcor2}
  \langle \e^{\i 2 \pi {\vec{G}} \cdot {\vec{h}}({\vec{r}})} 
    \e^{-\i 2 \pi \vec{G} \cdot \vec{h}(0)} \rangle = 
     \e^{-\frac{1}{2}
     \langle [2\pi\vec{G}\cdot (\vec{h}(\vec{r}) - \vec{h}(0))]^2 \rangle} \; ,
\ee 
and the height-height correlation function calculated in the Gaussian field
theory defined by \eqref{freegauss} (for $r\gg a$; $a$ is  the lattice spacing)
\be{3hvar}
 \langle (h_i(\vec{r}) - h_j(0))^2 \rangle = {\rm const} +  
 \frac{\delta_{ij}}{2 \pi^2 g} \ln|\vec{r}| \; ,
\ee
we find:
\be{3Eldim}
x(\vec{G}) = \frac{1}{2g} \: |\vec{G}|^2 \; .
\ee

\begin{figure}[tb]
\epsfxsize=7cm \epsfbox{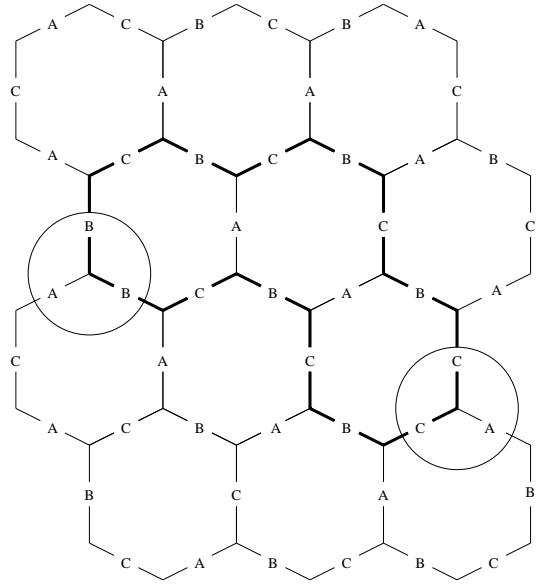}
\caption[A vortex-antivortex pair in the three-colouring model.]
{
\label{3coldefectsfig}
Elementary defects in the three-colouring model are associated with loops 
of alternating colour;  
exchanging the two colours (B and C) 
along one half of the  loop (shown in bold) will generate defects (circled), 
that is violations of the 
edge colouring constraint, at the two ends. In the interface representation
these defects become vortex-antivortex configurations of the height.
}
\end{figure}

Operators with a non-zero magnetic charge $\vec{b} \in {\cal R}$ can be 
associated with a vortex configuration of the height field, or a violation 
of the edge colouring constraint in the  
three-colouring model; see \fref{3coldefectsfig}. 
The topological charge of the vortex is given by $\vec{b}$,  
i.e., the height mismatch around a vertex of the honeycomb lattice,
and it can be calculated using 
the height rule introduced in \sref{effth}.

%----------------------------------------------------------------------------

%---------------------------------------------------------------------------

For a magnetic-type operator the scaling dimension $x(\vec{b})$
follows from the 
expression for the (dimensionless) 
interaction energy of a vortex--antivortex pair \cite{Xnienrev}
(for $r\gg a$)
\be{3vorint}
E_{\rm int} =  g |\vec{b}|^2 \ln{r} + {\rm const}  \; .
\ee
Here $\pm\vec{b}$ are the topological charges
of the vortices and $r$ is the distance between them; 
the above  expression for $E_{\rm int}$ 
follows from the  Gaussian form of the free energy, \eqref{freegauss}, and
is simply the two-dimensional version of Coulomb's law.   
The vortex-antivortex correlation function 
is given by the Boltzmann factor 
\be{bolzman}
\e^{-E_{\rm int}} \sim \frac{1}{r^{2x(\vec{b})}} \; ,
\ee
where $x(\vec{b})$ is fixed by  the coupling $g$,
\be{3vordim}
  x(\vec{b}) = \frac{g}{2} \: |\vec{b}|^2 \; .
\ee  

Equations \eref{3Eldim} and \eref{3vordim} specify the complete 
spectrum of scaling dimensions in the $n=2$ FPL model. Next we 
turn to the problem of {\em identifying} specific correlation
functions in this model associated with these exponents. 

The magnetic dimension $x_{\rm h}$, introduced by Bl\"ote and Nienhuis 
\cite{Xblote}, governs the
probability that two points on the honeycomb lattice lie on the same
loop of the FPL model. The loops  in the $n=2$ FPL model are  
{\em contour} loops of a particular component
of the height field. This is an important observation
which will lead us to the exact value of the coupling $g$, 
and we clarify it further. 

Say we choose the $BC$ loops in the three colouring 
model to represent the loops in the FPL model. Now  
consider the points at the centers of the hexagonal 
plaquettes along one side 
of a $BC$ loop. 
These points are separated by $A$ coloured bonds, and
consequently the component of the microscopic height $\vec{z}$
orthogonal to $\vec{A}$ is unchanged as we go along the
loop. Recently it has been argued that 
the exponent associated with the correlation function that measures the 
probability that two points belong to the same contour loop 
of a random Gaussian
surface is independent of the stiffness ($2\pi g$), 
and equal to $1/2$ \cite{XKHPRL}.
Therefore, we conclude, the magnetic dimension in the $n=2$ FPL 
model is $x_{\rm h}=1/2$.  The numerical result
of Bl\"ote and Nienhuis is $x_{\rm h}=0.470(1)$ (see Table I in \cite{Xblote}), 
and $x_{\rm h}=1/2$ was also found from the Bethe Ansatz solution \cite{Xbatch}.

In the 
interface representation of the three-colouring model the magnetic 
dimension can also be 
associated with a vortex-antivortex correlation function with 
the magnetic vector-charge
\be{b1}
\vec{b}_h = \vec{B}-\vec{C} = (0, 1) \; .
\ee
This comes about in the 
following way.  
The correlation function that measures the probability that two points 
separated by $\vec{r}$
belong to  the same $BC$ loop is $Z(\vec{r})/Z_0$.
The restricted partition function $Z(\vec{r})$ 
is simply the number of colourings with a $BC$ loop 
passing through $\vec{0}$ and $\vec{r}$, 
while $Z_0$ is the total number 
of colourings. 
Now, if we exchange the two
colours on the loop along one half of the loop going from $\vec{0}$ to 
$\vec{r}$, 
then this will  generate a vortex and an antivortex with charges 
$\pm\vec{b}_{\rm h}$, at these two points.
%% see \fref{3coldefectsfig}. 
The magnetic dimension $x_{\rm h}$ is therefore
\be{xhxb}
x_{\rm h} = x(\vec{b}_{\rm h}) \; .
\ee
Earlier we found $x_{\rm h}=1/2$, 
and from \eqref{3vordim} we can now calculate the {\em exact} value of the
coupling in the $n=2$ FPL model: 
\be{gnum}
 g = 1 \; .
\ee

Huse and Rutenberg \cite{Xhuse} deduced the value of the coupling
constant $g$ for $n=2$ from the exact solution of the three-colouring 
model due to Baxter
\cite{Xbaxter}. They showed that from Baxter's solution one can conclude 
that the interface  model is {\em exactly} at its roughening
transition. In standard scaling terms this means that the dimension of 
the locking potential, $V(\vec{h})$ in \eqref{free1}, is 2. This on the 
other hand leads to the
equation $x(\vec{G}_{\rm V})=2$, where $\vec{G}_{\rm V}$ is the smallest 
electric vector-charge appearing in the Fourier expansion 
\eref{fourier} of the operator $V(\vec{h})$; $\vec{G}_{\rm V}$ is also 
the second 
shortest vector in $\cal{R}^\ast$. From \eqref{3Eldim} and    
$|\vec{G}_{\rm V}|=2$ the value of the coupling, $g=1$, follows.   

Another dimension that was measured by Bl\"ote and Nienhuis is the 
{\em temperature dimension}  which is associated with a vertex of the 
honeycomb lattice not covered by a loop. 
An uncovered vertex becomes in the three-colouring model a defect with 
the same colour on all three of its surrounding bonds \cite{XKH};
in the interface representation  this becomes
a vortex with a topological charge\footnote{Here and throughout we
take the loops in the FPL model to be $BC$ coloured. Of course any 
other choice of colours would lead to the same results.}
\be{fplbt}
\vec{b}_{\rm t} = 3 \vec{A} \; ,
\ee
which is the second shortest  vector in $\cal R$. 
Hence, using \eqref{3vordim} for the dimension of a magnetic-type operator, 
and \eqref{gnum} for the value of the coupling, 
the temperature dimension is
\be{fpltempdim}
x_{\rm t} = x(\vec{b}_{\rm t}) = 3/2 \; .
\ee 
For comparison, 
the numerical transfer matrix result is $x_{\rm t}=1.46(1)$ (see 
Table I in \cite{Xblote}), and it
is in good agreement with our exact result when systematic errors are 
taken into account \cite{Xblote}. This dimension was not 
calculated  by  Batchelor {\em et al.} \cite{Xbatch}.

One possible
generalization of the loop correlation function is the so called 
{\em watermelon} correlation function 
 which measures the probability of having 
$m$ loop segments meeting at two point separated by $\vec{r}$ \cite{Xdupl};
$m=2$ would then be the loop correlation function. 
The central result of the Bethe Ansatz solution of the FPL model by 
Batchelor {\em et al.} 
\cite{Xbatch} was the calculation of the watermelon dimensions $x_m$. 
Here we show how they can be calculated from the Coulomb gas.  

In a way that is completely analogous to the above analysis of the loop 
correlation function,
the watermelon correlator becomes a vortex-antivortex correlation function.
Therefore, in order to 
calculate $x_m$ we need to determine the appropriate 
magnetic vector-charges $\vec{b}_m$. 
We divide 
this calculation up into four steps:
\begin{enumerate}
\item $m=1$ -- This corresponds to having a single $BC$ loop segment between 
points $\vec{0}$ and $\vec{r}$, which in turn implies 
that the colour configuration 
at $\vec{0}$ is $\{A,A,B\}$. Using the height rule around $\vec{0}$ 
we find  
\be{fplb1}
       \vec{b}_1 = 2 \vec{A} + \vec{B} = (\frac{\sqrt{3}}{2}, \frac{1}{2}) 
        \; .
\ee
\item $m=2$ -- In this case there is a $BC$ loop  
originating at 
$\vec{0}$ and ending at $\vec{r}$. The associated topological 
charge  was found earlier, \eqref{b1},  
\be{fplb2}
       \vec{b}_2 = \vec{b}_h = (0,1)  
\ee
\item $m=2k$ -- Here we have $k$ $BC$ loops
originating in the neighborhood of $\vec{0}$ 
and ending in the neighborhood of
$\vec{r}$. The total topological charge around the endpoints is
\be{fplb2k}
        \vec{b}_{2k} = k \vec{b}_2  \; .
\ee
%where ${\bf b}_2$ in \eqref{fplb2} is the Burgers vector associated with one
%$A$$C$ loop.
\item $m=2k-1$ -- This case can be thought of as having $k-1$  $BC$ 
loops and an additional $BC$ segment, originate
in the neighborhood of $\vec{0}$, and end in the neighborhood of
$\vec{r}$.
The total topological charge is
\be{fplb2k1}
          \vec{b}_{2k-1} = (k-1) \vec{b}_2  + \vec{b}_1 \; ,
\ee
where $\vec{b}_1$ and $\vec{b}_2$ are given by \eref{fplb1}
and \eref{fplb2}, respectively.
\end{enumerate}

Using the exact value of the coupling, $g=1$, and the above 
calculated magnetic vector-charges, from \eqref{3vordim} 
we find for the watermelon dimensions:
\bea{fplwaterdim}
    x_{2k} & = & \frac{1}{2} k^2  \nonumber \\ 
    x_{2k-1} & = & \frac{1}{2} (k^2 - k + 1) \; .
\eea

These exponents form a special case of the more general spectrum
of exponents associated with all the possible magnetic vector 
charges in the repeat lattice 
$\vec{b}_{j,k}=j(\vec{A}-\vec{B})+k(\vec{A}-\vec{C}) 
\in {\cal R}$
\be{exptable}  
x_{j,k} = x(\vec{b}_{j,k}) = \frac{1}{2}\left( j^2 + k^2 + j k \right) \; ,
\ee
where we have once again made use of \eqref{3vordim}.

These charges are associated with defects in the FPL model,
in which the full--packing constraint is violated locally.
Each defect can be contained in a non-selfintersecting 
polygon on the triangular lattice, dual to the original honeycomb lattice. 
The edges of the honeycomb lattice cut by the polygon will be called 
{\em defect edges}.

The topological charge of a defect in the solid-on-solid version of the 
model is the vector sum of the height differences measured along the 
polygon in, say the clockwise direction. This trivial procedure yields 
via \eqref{exptable} the exponent governing the spatial decay of 
correlations between two oppositely charged defects.
In the three-colouring model the charge of a defect is simply that of
the corresponding height configuration.

In the loop version of the model two-defect correlations are defined by 
the requirement that specified defect edges of the two defects be 
connected by a loop segment, and that the remaining edges be empty. 
Like everywhere else in the lattice the empty edges correspond to the colour 
$A$ and the occupied ones to $B$ or $C$. 
The possibility of a loop segment connecting defect  edges 
of the same defect must be excluded.
This can be done by maximizing the charge of the defect
using the choice between $B$ and $C$. 
Thus the corresponding topological charge of the one defect can be found by 
associating the vector $\vec{A}$ or $-\vec{A}$ to the empty defect edges and 
$\vec{B}$ or $-\vec{C}$ to the occupied ones. For the other defect the same 
rule applies with $\vec{B}$ and $-\vec{C}$ replaced by $\vec{C}$ and $-\vec{B}$.

It may be noted that the two sublattices of the honeycomb lattice are 
not equivalent when positioning a defect. For instance according to the 
above rule the charge of a vertex with three empty edges is $3\vec{A}$ on 
the one sublattice but $-3\vec{A}$ on the other. Two such defects have a 
total charge zero only if they are placed on different sublattices.
With periodic boundary conditions correlations functions between two 
defects that do not have opposite charge are zero. However, with open 
boundary conditions there are non-zero two-point correlations between 
different magnetic defects; for defect charges
$\vec{b}$ and $\vec{b'}$  
the exponent is $x(\vec{b}) + x(\vec{b}) - x(\vec{b}+\vec{b'})$.

\subsection{FPL model for  $n<2$}
\label{n<2}

Here we extend the Coulomb gas description of the FPL model for loop 
fugacities $n<2$. We show that this can be accomplished by introducing
a background vector-charge. The effect of the background charge is a 
lowering of the conformal charge, and a shift in the scaling dimensions
found above.  

Let us first introduce, in the three-colouring model, the staggered chirality
$\chi(\vec{r})$.
This operator takes two values: it is $+1$ ($-1$)  if
the colours go clockwise around the vertex $\vec{r}$ on the even (odd)
sublattice of the honeycomb lattice,  
and $-1$ ($+1$) if the colours go anti-clockwise. 
For $n<2$ the free energy \eref{freegauss} has an
extra imaginary bulk term conjugate to the 
$\chi(\vec{r})$; this term was introduced in the 
colouring model by Baxter \cite{Xbaxter}. 
The  effect
of the bulk term $\i \lambda \int \d^2\vec{r} \: \chi(\vec{r})$, 
is to assign a phase factor $\exp ({\pm \i \lambda})$ 
every time a loop in the FPL model  makes a left or a
right turn. This will have the effect of assigning to each loop a weight
\be{fplnlam}
    n = \e^{-\i 6 \lambda} + \e^{\i 6 \lambda} = 2 \cos(6\lambda) \; ,
\ee
due to the fact that the difference between the number of 
left and right turns, when walking  along  a 
closed loop on the honeycomb lattice, is six.
By inspection of the ideal 
state graph in \fref{3-ideal} we conclude 
 that the most relevant vertex operator  
appearing in the Fourier expansion of $\chi(\vec{r})$ has an electric 
vector-charge $\vec{G}_{\chi}$, which is the second largest vector in 
${\cal R}^\ast$ (same as for the locking potential). Therefore at
fugacity $n=2$
the scaling dimension of the chirality is $2$, i.e. it is marginal. We
expect that 
its only effect on the Gaussian action $f$ in \eqref{freegauss} 
will be an isotropic change  of the coupling $g$.\footnote{Here we
have assumed that the staggered chirality is an integrably marginal
operator \cite{Xchoud}. This, as will be shown later, is confirmed 
by Baxter's solution \cite{Xbaxter}.}   

Other then the marginal term, the shift of the loop fugacity 
away from $n=2$ will also generate a term in the free energy 
(Euclidean action) 
which couples the height field to the scalar
curvature \cite{Xfoda}. This can be
most readily understood by taking  
the  FPL model to be defined on a cylinder of circumference $L$. 
Namely, a seam running along the length of the cylinder  has to 
be introduced in order to give 
the correct fugacity $(n)$ for loops winding around the 
cylinder. A bond of colour $\vec{\sigma}\in\{\vec{A},\vec{B},\vec{C}\}$ 
which crosses the seam gets an extra factor
     $ \exp(\i 2 \pi \vec{E}_0 \cdot \vec{\sigma}) $ 
where the vector-charge 
\be{fplbckch}
-2\vec{E}_0=(0,-2e_0)
\ee
is the background charge in the Coulomb gas \cite{Xdots}.  
The value of $\vec{E}_0$ is chosen in such a 
way so that only the $B$ and $C$ bonds acquire a phase ($\pm\pi e_0$) 
when crossing the seam.
As a reminder, we
note that the 
$BC$ loops in the three-colouring model were chosen earlier to represent the 
loops of the FPL model; this choice is of course arbitrary, and any one of the
three possible choices of colour pairs would give the same results. 
Summing over the two possible ways of colouring a $BC$ loop, we find for the 
fugacity of loops winding around the cylinder:
\be{fplchfug}
    n =  2 \cos(\pi e_0) \; ,
\ee
which when compared to \eqref{fplnlam} gives the relation
\be{fplchfugII}
\pi e_0 = 6\lambda \; .
\ee
The effective field theory in the presence of the seam can be written as
\cite{XcardyaffPRL}:
\begin{equation} 
\label{contHam}
  f\:=\:\int \d^2\vec{r}\,\pi g(|\nabla h_1|^2+|\nabla h_2|^2)+2\pi
  \i e_0 \left(h_2(L,\infty)-(h_2(L,-\infty)\right). 
\end{equation}
This modification of the Gaussian field theory results in a shift of the
conformal charge,
\be{confch}
 c = 2 - 6 \frac{e_0^2}{g} \; ,
\ee
and it also has an effect on the scaling dimensions of operators. 
As shown by Dotsenko and Fateev \cite{Xdots}, the correlation
functions of the modified Coulomb gas can be expressed in terms 
of correlation functions in the Gaussian model \eref{freegauss}.
The non-zero two-point functions in the modified theory are
those of vertex operators having opposite electro-magnetic charge in
combination with a floating charge $2\vec{E}_0$ that precisely cancels
the background charge $-2\vec{E}_0$. The floating charge may combine
with the negative charge. An operator whose total electro-magnetic
vector-charge is $(\vec{G},\vec{b})$ then has a dimension
$x(\vec{G},\vec{b})$ given by \cite{Xdots} 
\begin{equation}
\label{elmagdim}
 x(\vec{G},\vec{b}\,)\;=\;\frac{1}{2g}\vec{G}\cdot(\vec{G}-2\vec{E}_0)+\frac{g}{2}|\vec{b}|^2 \; .
\end{equation}
Instead, the floating charge may split up so that both the positive
and negative charge get increase by $\vec{E}_0$. The dimension of such
an operator is given by $x(\vec{G}+\vec{E_0},\vec{b}\,)$.
This is the {\em complete} spectrum of scaling dimensions in the FPL
model, where $\vec{G}\in {\cal R}^\ast $, $\vec{b}\in {\cal R}$. The 
background charge is related to the fugacity by \eqref{fplchfug}, while 
the relation between the coupling $g$ and the fugacity $n$
remains unknown. We turn to this problem next. 

In the $n=2$ FPL model the staggered chirality 
(or equivalently, the locking potential) 
is marginal. It follows from the exact solution obtained by Baxter
\cite{Xbaxter} that it remains marginal as long as $\exp (\i 6\lambda)$
lies on the unit circle, i.e. $\lambda$ is real. Therefore, the
chirality is marginal in the entire
regime $0\leq n\leq 2$, and one expects it to change the value of the
renormalized coupling constant $g$ continuously with $n$.
The dimension of the staggered chirality, for $n<2$, is governed by the
second smallest vector in ${\cal R}^\ast$ 
{\em parallel} to $\vec{E}_0$. Using \eqref{elmagdim} with 
$\vec{G}_{\chi}=(0,2)$,  the
marginality of $\chi(\vec{r})$ (i.e., $x(\vec{G}_{\chi})=2$)
gives the relation between $g$ and $e_0$:
\begin{equation}
\label{e0vsg}
  \frac{4-4e_0}{2g}=2 \; ,
\end{equation}
where, from \eqref{fplchfug} we can read of the dependence of the background
charge on the fugacity,
\be{e0vsn}
 e_0 = 1 - g = \frac{1}{\pi} \arccos{\left(\frac{n}{2}\right)} \; .
\ee 
This agrees with the relation obtained numerically by Bl\"ote and
Nienhuis \cite{Xblote}, and the Bethe Ansatz result of Batchelor {\em et al.}

Equipped with equations~\eref{elmagdim}, \eref{e0vsg}, and \eref{e0vsn}
we can now calculate the 
watermelon dimensions $x_m$ for the $n<2$ FPL model. 
In the Coulomb gas representation
the watermelon scaling dimensions
are given by $x(\vec{E}_0, \vec{b}_m)$, 
where the magnetic charge is given by \eqref{fplb2k} for $m$ even, 
and by \eqref{fplb2k1} for $m$ odd. 
The electric charge $\vec{E}_0$ 
is due to the electric-type operators 
$\exp(\i 2 \pi \vec{E}_0 \cdot \vec{h})$ that must be inserted  
at the endpoints of 
the watermelon configuration in order to correct for the spurious  phase
factors $\exp(\pm \i 6\lambda)$
that arise due to the winding of the loop segments 
around the endpoints \cite{Xnienrev}.  
We conclude that the watermelon scaling dimensions are given by
%the watermelon dimensions follow from:
\bea{FPLwaterdims}
 x_{2k} = x(\vec{E}_0, \vec{b}_{2k})
                 & = & \frac{g}{2} k^2 - \frac{(1-g)^2}{2g} \nonumber \\
 x_{2k-1} = x(\vec{E}_0, \vec{b}_{2k-1})
                 & = & \frac{g}{2} (k^2 - k + 1) -  \frac{(1-g)^2}{2g} \; ,
\eea    
which was also found by Batchelor {\em et al.}, (see equations (16) and (17) 
in \cite{Xbatch}) and it generalizes the $n=2$ result \eref{fplwaterdim}. 
The dimension $x_2$ was also calculated numerically by 
Bl\"ote and
Nienhuis \cite{Xblote} for different values of $n$, 
and their results agree very well with the exact values.

The temperature dimension $x_{\rm t}$, 
for different values of $n$, was also determined
numerically by Bl\"ote and Nienhuis, 
but this dimension does not appear in 
the Bethe Ansatz solution of Batchelor {\em et al.}. 
The dimension $x_{\rm t}$ is
related to a defect in the FPL model associated with an uncovered vertex. 
Unlike the case of the watermelon dimensions
 there are no loop segments associated with this defect
%as in the case of the watermelon dimensions
and there is consequently 
no need for an electric type operator to correct for the winding of the loop
segments around the endpoints. Therefore, we have
\be{fpltempdim2}
x_{\rm t} = x(0,\vec{b}_{\rm t}) =  \frac{3}{2} g \; ,
\ee
where the magnetic vector-charge 
$\vec{b}_{\rm t}=3\vec{A}$ (\eref{fplbt}).
The numerical results
in \cite{Xblote} are in very good agreement with our exact result.

Once again the equations \eref{FPLwaterdims} and \eref{fpltempdim2} 
are special cases of a more general spectrum 
\be{jknew} 
 x_{j,k} = \frac{g}{2}\left( j^2 + k^2 + j k \right) - \frac{(1-g)^2}{2 g} 
 \left(1- \delta_{j,k}\right) 
\ee
for the charge  $\vec{b}_{j,k} = 
j (\vec{A}-\vec{B})+k(\vec{A}-\vec{C})$.
These, like \eqref{exptable}, govern the correlations between mixed 
defects defined by empty and occupied defect edges and the requirement 
that the occupied edges of both defects are mutually connected.  
Only for defects with topological 
charges purely in the $\vec{A}$ direction is  the exponent 
unaffected by the background charge.

Another way of viewing the  effect that the background
charge has on the exponents is the following. Consider the transfer
matrix of the FPL model on a cylinder of circumference $L$. In terms of the
eigenvalues of the transfer matrix 
($\lambda_0 > \lambda_1 > \lambda_2 > \ldots$)
the exponents are defined by the 
gaps between the largest eigenvalue and the smaller ones
\cite{XcardyaffPRL},
\begin{equation} 
  x_i\;=\;\frac{L}{2\pi}\ln{\frac{\lambda_0}{\lambda_i}} \; .
  \label{gaps}
\end{equation}
The free energy of the FPL model 
is given by the logarithm of the largest eigenvalue, 
and its finite size scaling is given by
\begin{equation}
f_L\;=\;\frac{1}{L}\ln{\lambda_0}\;\simeq\;f_{\infty}+\frac{\pi
  c}{6 L^2} \; ,
\end{equation}
where $f_{\infty}$ is the bulk free energy, and $c$ the central 
charge.
The background charge modifies the ground state of the system
through the central charge, \eqref{confch}, and therefore 
the various exponents, corresponding to
excited states with respect to this shifted ground state, will also change. 
In the case of the watermelon dimensions there are extra lines
running along the cylinder. 
This means that for this
excited state we should  ``turn off'' the seam, otherwise one would count
the winding number of the loops around the cylinder which are not
permitted, since they would intersect the lines along the cylinder.
This state will 
therefore not have the correction from the background charge. As it is
measured with respect to the new ground state via \eqref{gaps}, the
watermelon dimension $x_m$ will not only differ from the $n=2$ value 
($x(\vec{b}_m)$, \eqref{fplwaterdim})
due to the  different  value of
$g$, but also by a shift, which is simply $1/12$ of the shift of the 
central charge in \eqref{confch}. 
Therefore, the  value of the watermelon dimensions is 
\begin{equation} 
  x_m\;=\; x(\vec{b}_m) - \frac{6 (1-g)^2}{12 g} 
\label{xh}
\end{equation} 
which coincides with \eqref{FPLwaterdims}.
In the state corresponding to the thermal excitation  
loops going around the cylinder are permitted 
and we need the seam to count those
contributions correctly. This state therefore gets
shifted in the same way as the ground state. The thermal exponent
therefore only differs from the $n=2$ exponent in the value of $g$, and
is given by \eqref{fpltempdim2}. 

\section{Geometrical Properties of Loops}
\label{geom}

In this section we consider the geometrical properties 
of loops in the FPL model. We calculate the fractal dimension of loops
and the loop length  distribution function.

The length of a loop $s$, and its radius $R$ are related by
\begin{equation}
\label{fdim}
s \sim R^{D_{\rm f}} \; ,
\end{equation}
where $D_{\rm f}$ is the fractal dimension of the loop. The loop radius is defined as
the radius  of the smallest circle that contains the loop.

The distribution of loop lengths $P(s)$ measures the probability
that a loop, in a fully packed configuration, passing through a chosen 
vertex
has length $s$. It is given by a power law:
\begin{equation}
\label{taudef}
 P(s) \sim s^{-(\tau - 1)} \; .
\end{equation}

The geometrical exponents
$D_{\rm f}$ and $\tau$ can be related by a scaling argument 
to the exponent $x_2$ \cite{Xsaleur,Xcardy} 
\begin{equation}
\label{relations}
D_{\rm f}  = 2 - x_2 \; , \; \; \; \; \tau - 1 = \frac{2}{2 - x_2} \; .
\end{equation}
Using the calculated value of $x_2$, \eqref{FPLwaterdims}, we find:
\begin{equation}
\label{relations2}
D_{\rm f} = 1 + \frac{1}{2g} \; , \; \; \; \; \tau - 1 = \frac{4g}{1+2g} \;.
\end{equation}

In the limit $n\rightarrow 0$ ($g=1/2$) we find $D_{\rm f} = 2$, which is what
we expect for Hamiltonian walks. For $n=2$ we find $D_{\rm f} = 3/2$ which is
the fractal dimension of equal height (contour) loops on a random
Gaussian surface \cite{XKHPRL}. Indeed, for $n=2$ the background charge is
equal to zero and the free energy $f$ defines a random Gaussian
surface; the loops in the FPL model are contour loops of this random
surface. 

In the Kagom\'e Potts model representation of the $n=2$ FPL model
the exponent $\tau - 1$ has been determined  numerically to be
$1.34 \pm 0.02$ \cite{Xchandra}, in good agreement with the exact result
$\tau - 1 = 4/3$ which follows from \eqref{relations2} for $g=1$. 

\section{Summary and remarks}

We have calculated the conformal charge and the exact exponents in the 
fully packed loop model on the honeycomb lattice. To this end we 
proposed a simple conformal field theory for the scaling limit of 
this model, and it is given by the vacuum phase of a two dimensional
Coulomb gas with an added background charge. The magnetic and electric 
charges of the Coulomb gas 
were found to be {\em vectors} in the triangular lattice
${\cal R}$ (the so-called ``repeat'' lattice) and its dual
${\cal R}^\ast$. These charges give the complete operator spectrum 
of the fully packed loop model of which the watermelon dimensions 
calculated by Batchelor {\em et al.} are a subset. The exact value of the 
temperature dimension found here and not in the Bethe Ansatz solution
are in agreement with the numerical results of Bl\"ote and Nienhuis
\cite{Xblote}.

Coulomb gas methods with vector-charges have been used
previously by Fateev and Zamolodchikov \cite{Xfatzam} 
to calculate correlation functions in
the $\Zs_3$ models. Their work was extended  by Pasquier \cite{Xpasq}  who
considered the continuum limit of lattice models with quantum group symmetries.
The FPL model is most likely related to the models of  Fateev and 
Zamolodchikov since in the $n\rightarrow 2$ limit the FPL model 
has an enlarged chiral symmetry,  
given by the $su(3)_{k=1}$ Kac-Moody algebra \cite{XKHNPB}. This is also true
of the  $\Zs_3$ models. Moreover, for $n<2$ we  introduced a 
term proportional to the staggered chirality into the free-energy (action) of
the FPL model. The presence of this term breaks the full permutation symmetry
of the three-colouring model down to $\Zs_3$; only cyclic permutations of the
colours leave the staggered chirality unchanged.    

Finally,  is interesting to note that in the folding model 
the microscopic heights $\vec{z}(\vec{r})$  
specify the {\em positions} of the vertices,  $\vec{r}$, 
of the triangular lattice in the folded state. The coarse grained (entropic
in origin) free energy, 
corresponding to 
the different ways of folding the triangular lattice, is  
Gaussian, \eqref{freegauss}.
In the theoretical considerations of tethered membranes this is
usually the starting assumption which is  justified by the results of
numerical simulations \cite{XKKN}. 
We see that in this simple folding model the 
gradient squared form of the entropic contribution to the free energy of 
folding, is closely related to the
fact that the FPL  model maps to the vacuum  phase of a
two-dimensional Coulomb gas.

The ideal states we had identified in the three colouring model are the states
that are entropically selected. They are flat, $\vec{h}={\rm const}$, 
and in the folding model they map to 
states in which the whole triangular lattice has been folded down to a 
single triangle. Introducing an energy penalty associated with folding might
stabilize a flat state of the membrane at low temperatures, 
while the entropically selected
folded states would necessarily win at high temperatures. Therefore, we might 
expect
a {\em folding transition} to occur at some intermediate temperature. Such a 
transition was
found in the numerical transfer matrix calculation of Di Francesco and
Guitter \cite{Xdifrancesco}. It would be interesting to study this
transition using the Coulomb gas methods developed here. It is our 
hope that  this will allow us to calculate properties of this intriguing
transition exactly.  

\ack
We wish to thank C.L. Henley and J. Cardy for discussions and C.L. Henley
for a careful reading of the manuscript.

\end{document}